\documentclass{moriond}

\def\Journal#1#2#3#4{{#1} {\bf #2}, #3 (#4)}

\def\PRD{{\em Phys. Rev.} D}

\def\be{\begin{equation}}
\def\ee{\end{equation}}
\def\bea{\begin{eqnarray}}
\def\eea{\end{eqnarray}}

\newcommand{\Photo}{\includegraphics[height=35mm]{mypicture}}

\begin{document}
\vspace*{4cm}
\title{Search for Non-Standard Interactions by atmospheric neutrino}

\author{ S. Fukasawa }
\author{ \it Department of Physics, Tokyo Metropolitan University, Minami-Osawa,\\
 Hachioji, Tokyo 192-0397, Japan}

\maketitle\abstracts{
We investigate the effects of neutral current Non-Standard Interactions in
propagation on atmospheric neutrino experiments such as Super-Kamiokande
and Hyper-Kamiokande.  With the ansatz where the parameters which have
strong constraints from other experiments are neglected, we show how these
experiments put constraints on the remaining parameters of the Non-Standard
Interactions.}

\section{Introduction}
Neutrinos are the least tested particles of the Standard Model
(SM). Furthermore non-zero neutrino masses, which cannot be explained
by the SM, have been measured. Non-zero neutrino masses cause neutrino
oscillation which may play an important role in searching for physics
Beyond the Standard Model (BSM). All the three mixing angels in the
standard three flavor neutrino oscillation have been measured by
2012. Tasks left for us are to measure the Dirac CP phase and the sign of
$\Delta m^2_{31}$. It is expected that future high-intensity
long-baseline experiments (T2HK,LBNO,LBNE etc.) may be able to measure
these quantities and allow high precision measurement of the oscillation parameters. Hence the future experiments can take us to a new stage in which BSM is searched by looking at deviation from the SM with neutrino masses. For this reason, it is important to study new physics in neutrino sector.
\section{New Physics}
We consider flavor-dependent exotic couplings of neutrinos with matter in this talk. These Non-Standard Interactions (NSI) are divided into Charged-Current (CC) interactions and Neutral-Current(NC) interactions. We consider only non-standard NC interactions in propagation because constraints for non-standard CC interactions at detectors and sources are much stronger than those in propagation. They are expressed by effective 4-fermi interactions:
\begin{equation}
{\cal L}_{NP}=-2 {\sqrt{2}}G_{F}[\bar \nu_{\alpha} \gamma^{\mu}\nu_{\beta}][
\epsilon_{\alpha\beta}^{f_L}\bar f_L \gamma_\mu
f_L+\epsilon_{\alpha\beta}^{f_R}\bar f_R\gamma_\mu f_R],
\label{L}
\end{equation}
where $\alpha,\beta$ are neutrino flavor indices and $\epsilon_{\alpha\beta}^{f_{L,R}}$ denotes the strength of the NSI between the neutrinos $\nu$ and left-handed (right-handed) components of the fermions $f$. It is known that neutrinos feel potentials as matter effects when they are propagating in matter. If NSI exist, neutrinos feel additional matter effects and the oscillation probability is modified significantly. NSI can be parameterized in the form of matter effects
\begin{equation}
{\cal A}_{NSI}=A_{\rm CC}
\left(\begin{array}{ccc}
\epsilon_{ee} & \epsilon_{e\mu} & \epsilon_{e\tau} \\
\epsilon_{e\mu}^* & \epsilon_{\mu\mu} & \epsilon_{\mu\tau} \\
\epsilon_{e\tau}^* & \epsilon_{\mu\tau}^* & \epsilon_{\tau\tau} \\
\end{array}\right),
\end{equation}
where
\begin{equation}
\epsilon_{\alpha \beta}=\sum_{P=L,R}\sum_{f=e,u,d} \epsilon^{f_P}_{\alpha\beta}n_f/n_e .
\end{equation}
Here $n_f$ denotes number density of fermion $f$ and $A_{\rm CC}=\sqrt{2}G_F n_e$. In the case of atmospheric neutrino experiments, neutrinos go through the Earth, which is composed of protons($uud$), neutrons($udd$), and electrons and their number densities are approximately equal. Then we have
\begin{equation}
\epsilon_{\alpha\beta}\simeq\sum_{P=L,R}(\epsilon_{\alpha\beta}^{e_P}+3\epsilon_{\alpha\beta}^{u_P}+3\epsilon_{\alpha\beta}^{d_P}).
\label{epsilon}
\end{equation}
Matter effects affect the oscillation probability mainly when the
baseline is longer and the neutrino energy is in the range $10  {\rm GeV} < E < 100  {\rm GeV}$.  Therefore atmospheric neutrino experiments such as Super-Kamiokande(SK) and Hyper-Kamiokande(HK) are suited to search for NSI. 

\section{Constraints on NSI}
NSI are constrained by various experiments.  For instance the most
direct limits on  $\epsilon_{\alpha\beta}$ are given by experiments of
scattering of neutrino beams on target. This kind of experiments
give constraints on each $\epsilon_{\alpha\beta}^{f_{L,R}}$ in (\ref{L}), while the neutrino oscillation experiments can constrain the total sum $\epsilon_{\alpha\beta}$ in (\ref{epsilon}).

\subsection{Constraints from terrestrial experiments}
Constraints form terrestrial experiments are deduced as \cite{Biggio:2009nt} :
\begin{equation}
\left(\begin{array}{lll}
|\epsilon_{ee}|< 4\times 10^0 & |\epsilon_{e\mu}|<  3\times 10^{-1}& |\epsilon_{e\tau}|< 3\times 10^0 \\
 & |\epsilon_{\mu\mu}|< 7\times 10^{-2} & |\epsilon_{\mu\tau}|< 3\times 10^{-1} \\
 &  & |\epsilon_{\tau\tau}|< 2 \times 10^1 \\
\end{array}\right).
\end{equation}
The muon sector is strongly constrained, so we set $\epsilon_{e\mu},\epsilon_{\mu\mu},\epsilon_{\tau\mu}$ to zero in our analysis. On the other hand, the others are less constrained and hence there are rooms for improvement. 

\subsection{Constraints from the high energy behavior of atmospheric neutrino}
The high energy atmospheric neutrino data are well described by vacuum
oscillation between $\nu_{\mu}$ and $\nu_{\tau}$ but NSI can modify the oscillation probability significantly. Friedland-Lunardini \cite{Friedland:2004ah} pointed out that large NSI are consistent with the data, provided that epsilons have the following relation:
\begin{equation}
 \epsilon_{\tau \tau} \simeq \frac{|\epsilon_{e \tau}|^2}{1+\epsilon_{e e}}.
\end{equation}

\subsection{Summery of the constraints on NSI}
Taking these constraints on NSI into consideration, we analyze with the following ansatz:
\begin{equation}
\left(\begin{array}{ccc}
\epsilon_{ee} & \epsilon_{e\mu} & \epsilon_{e\tau} \\
\epsilon_{e\mu}^* & \epsilon_{\mu\mu} & \epsilon_{\mu\tau} \\
\epsilon_{e\tau}^* & \epsilon_{\mu\tau}^* & \epsilon_{\tau\tau} \\
\end{array}\right)
=
\left(\begin{array}{ccc}
\epsilon_{ee} & 0 & |\epsilon_{e\tau}|e^{i\phi} \\
0 & 0 & 0 \\
|\epsilon_{e\tau}|e^{-i\phi} & 0 & \frac{|\epsilon_{e\tau}|^2}{1+\epsilon_{ee}} \\
\end{array}\right). \
\end{equation}
Here $\phi$ is a phase of $\epsilon_{e\tau}$.
Furthermore we introduce the following ratio:
\begin{equation}
|\tan\beta| \equiv \frac{|\epsilon_{e\tau}|}{1+\epsilon_{ee}}.
\end{equation}
From the data of SK, Friedland-Lunardini \cite{Friedland:2005vy} gave constraints $|\tan \beta | <1.5$
at 2.5$\sigma$ C.L. \footnote{The constraints on NSI with large $\epsilon_{\alpha\beta}~(\alpha=e,\tau)$ from SK have also
  been given in
  Refs.~\cite{GonzalezGarcia:2011my,Mitsuka:2011ty,Gonzalez-Garcia:2013usa} with ansatz different from ours.}. What we want to do is to give an allowed region, which is obtained from SK or is expected from HK, in the $\epsilon_{ee}-|\epsilon_{e\tau}|$ plane.

\section{Results}
In our $\chi^2$ analysis, we fix the following oscillation parameters
\begin{equation}
\sin^2 2\theta_{12}=0.86, \hspace{0.4cm} \sin^2 2\theta_{13}=0.1,\hspace{0.4cm} \Delta m^2_{21}=7.6\times10^{-5} {\rm eV},
\end{equation}
while taking $\theta_{23}$, $\Delta m^2_{31}$,
$\delta_{CP}$ and $\phi$ as free.
To obtain a 2-dimensional allowed region,
$\chi^2$ is marginalized over $\theta_{23}$, $\Delta m^2_{31}$,
$\delta_{CP}$, $\phi$. In the case of SK, we look at
difference of significance from the SK data.  In the case of HK, we look at
difference of significance from the standard case which is predicted with
the standard oscillation parameters in our calculation code. The
numbers of events of atmospheric neutrinos are calculated based on
Ref. \cite{Yasuda:1998mh}.
\begin{figure}[h]
\centering
\includegraphics[width=5.8cm,angle=-90]{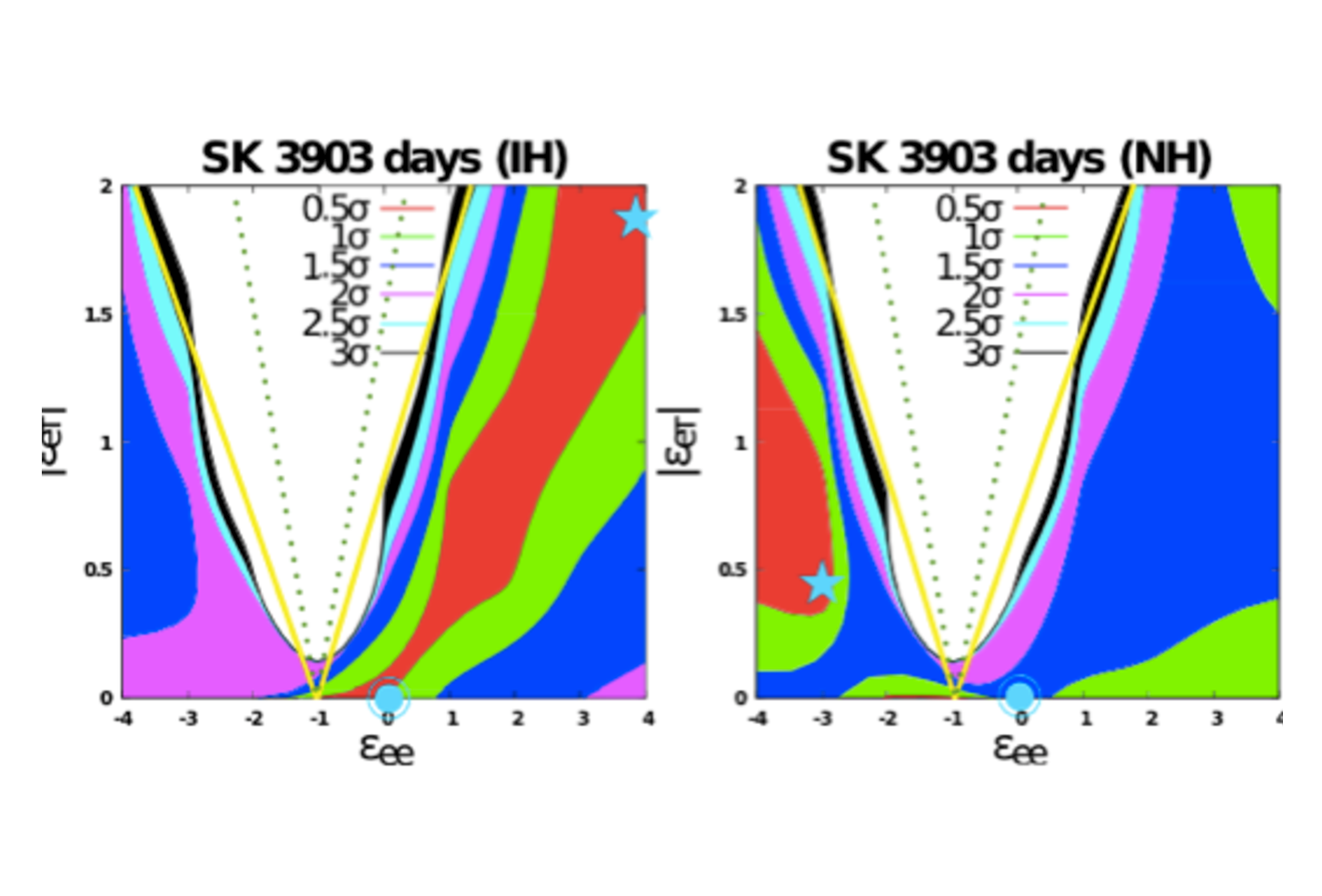}
\caption{Constraints on NSI at SK with NH(right) and IH(left). Dotted lines denote constraints on $|\tan\beta|<1.5$  given by Friedland-Lunardini in 2005. Solid lines are our results($|\tan\beta|<0.8$). These lines are drawn at $2.5\sigma$ C.L. Stars are best fit points and double circles denote standard case($\epsilon_{\alpha\beta}=0$). }
\label{sk}
\end{figure}

\begin{figure}[h]
\centering
\includegraphics[width=6cm,angle=-90]{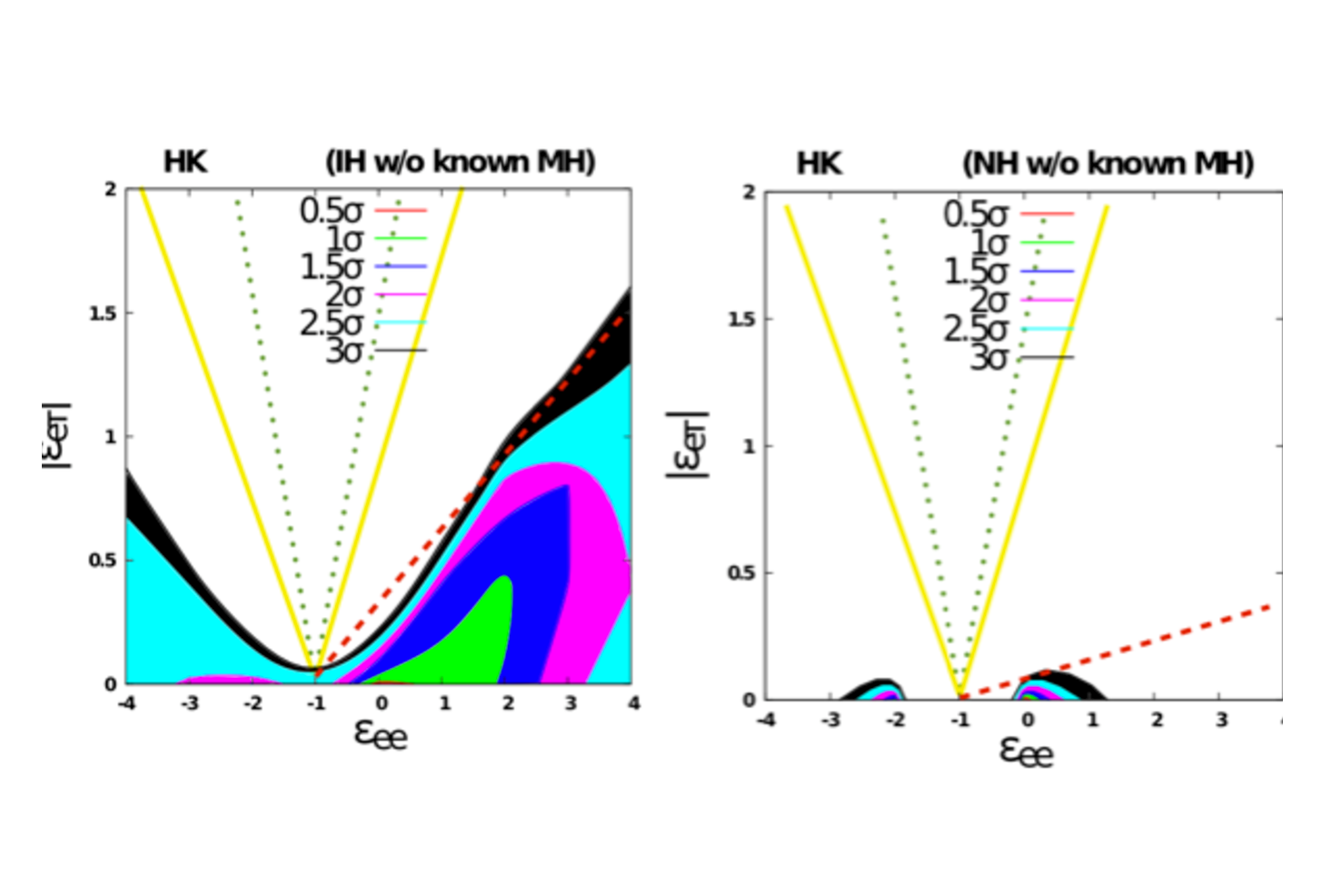}
\caption{Constraints on NSI at HK with NH(right) and IH(left). They are drawn without knowledge of mass hierarchy. The dotted and solid lines are as with fig.\ref{sk} and dashed lines are constraints on NSI at HK. In the case of NH($|\tan\beta|<0.06$) and IH($|\tan\beta|<0.3$).}
\label{hk}
\end{figure}

\begin{figure}[h]
\centering
\includegraphics[width=4.5cm,angle=-90]{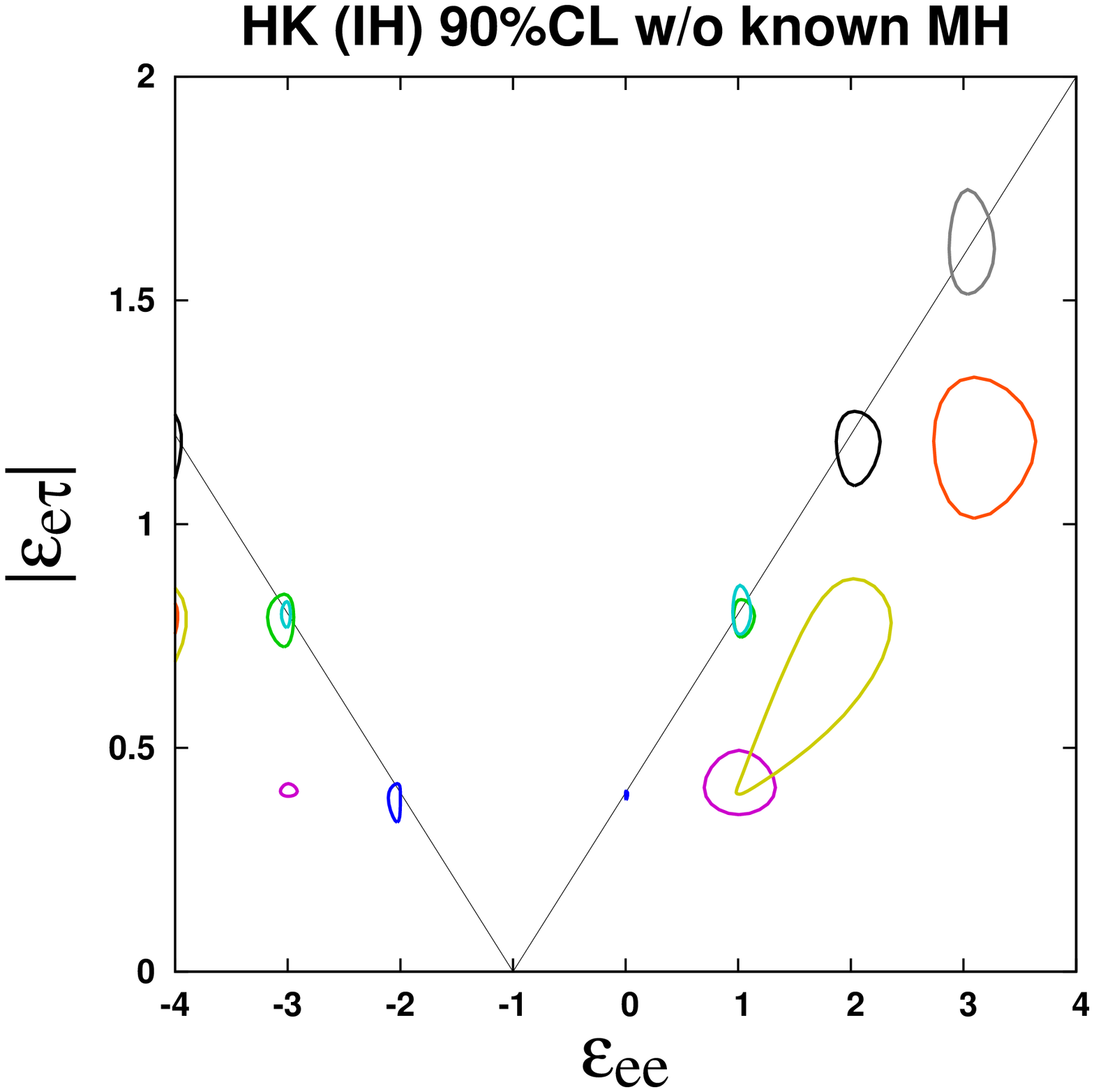}
\includegraphics[width=4.5cm,angle=-90]{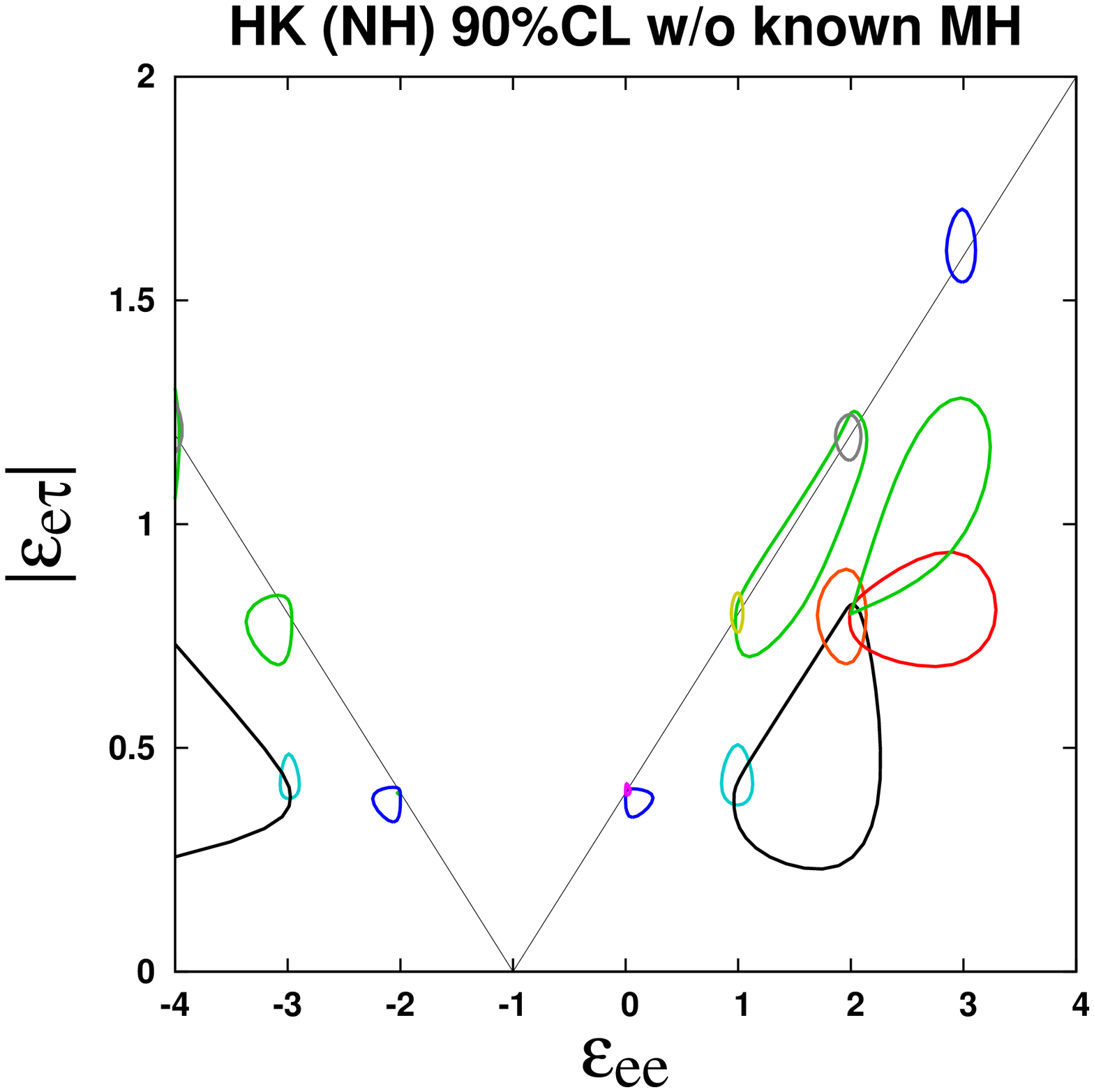}
\caption{Sensitivity to non-zero NSI parameters at HK(90\% C.L.) }
\label{sensitivity}
\end{figure}

Figure \ref{sk} shows constraints on NSI from SK. Although updated data
of SK (3903 days) are used, large NSI are not excluded. In addition,
the standard case is not the best fit. The reason that a scenario with new physics is preferred may be because we have not reproduced SK MC results completely. However, the excluded region is improved compared with the old one given by Friedland-Lunardini in 2005. Figure \ref{hk} shows constraints on NSI at HK.
The excluded region from HK is improved compared with SK and the
region $|\epsilon_{e\tau}| > 1 $ is excluded. Finally figure
\ref{sensitivity} shows sensitivity to the non-zero NSI parameters at
HK(90\% C.L.). In the presence of new physics, HK can determine NSI to
some extent with 90\% C.L.. The results at HK are first obtained in this work.

\section{Conclusion}
Taking into consideration the constraints from the terrestrial
experiments and the high energy behavior of atmospheric neutrinos,
i.e., with the ansatz :
$\epsilon_{e\mu}=\epsilon_{\mu\mu}=\epsilon_{\mu\tau}=0$ and
$\epsilon_{\tau\tau}=|\epsilon_{e\tau}|^2/(1+\epsilon_{ee})$, we have
searched for NSI. Under the ansatz we studied sensitivity to NIS of
$\nu_e - \nu_{\tau}$ sector in propagation at SK and HK.  Then the
excluded region at SK is improved compared with the old one given by
Friedland-Lunardini in 2005, because we used the updated SK data. The
excluded region expected at HK is obtained and HK is expected to improve constraints. In addition we studied sensitivity to non-zero NSI parameters at HK. If NSI are sufficiently large, HK can determine $\epsilon_{ee}$ and $|\epsilon_{e\tau}|$ to some extent.

\section*{Acknowledgments}
This work is in collaboration with Osamu Yasuda. I would like to thank the organizers for giving me an opportunity to present our work. 
This research was partly supported by a Grant-in-Aid for Scientific
Research of the Ministry of Education, Science and Culture, under Grant
No. 25105009.

\end{document}